\documentclass{aastex631}

\usepackage{bm}

\newcommand{\bec}[1]{\mbox{\boldmath $ #1$}}
\newcommand{\meanrho}{\overline{\rho}}

{}
{}
{}
\newcommand{\meanBB}{\overline{\mbox{\boldmath $B$}}{}}{}
{}
{}
{}
{}
{}
{}
{}
{}

{}
{}
{}
\newcommand{\meanA}{\overline{A}}
\newcommand{\meanB}{\overline{B}}

\def\half{{\textstyle{1\over2}}}

\begin{document}

\title{Forecast of solar activity based on mean-field dynamo model and neural network}

\author[0000-0002-5744-1160]{N. Kleeorin}
\affiliation{IZMIRAN, Troitsk, 108840 Moscow Region,  Russia}
\affiliation{Department of Mechanical Engineering, Ben-Gurion University of Negev, POB 653, 84105 Beer-Sheva, Israel}
\affiliation{Institute of Continuous Media Mechanics, Korolyov str. 1, Perm  614013, Russia}
\author[0000-0003-1677-4417]{K. Kuzanyan}
\affiliation{Institute of Continuous Media Mechanics, Korolyov str. 1, Perm  614013, Russia}
\affiliation{IZMIRAN, Troitsk, 108840 Moscow Region,  Russia}
\author{N. Safiullin}
\affiliation{Institute of Continuous Media Mechanics, Korolyov str. 1, Perm  614013, Russia}
\affiliation{N.N. Krasovskii Institute of Mathematics and Mechanics (IMM UB RAS), 620108 Ekaterinburg, Russia}
\author[0000-0001-7308-4768]{I. Rogachevskii}
\affiliation{Department of Mechanical Engineering, Ben-Gurion University of Negev, POB 653, 84105 Beer-Sheva, Israel}
\author[0000-0001-5100-806X]{V. Obridko}
\affiliation{Institute of Continuous Media Mechanics, Korolyov str. 1, Perm  614013, Russia}
\affiliation{IZMIRAN, Troitsk, 108840 Moscow Region,  Russia}
\author{S. Porshnev}
\affiliation{N.N. Krasovskii Institute of Mathematics and Mechanics (IMM UB RAS), 620108 Ekaterinburg, Russia}
\author[0000-0001-8098-0720]{R. Stepanov}
\affiliation{Institute of Continuous Media Mechanics, Korolyov str. 1, Perm  614013, Russia}

\begin{abstract}

We discuss a prediction of the solar activity on a short time-scale
applying the method based on a combination
of a nonlinear mean-field dynamo model and the artificial neural network.
The artificial neural network which serves as a correction scheme for the forecast,
uses the currently available observational data
(e.g., the 13 month running average of the observed
solar sunspot numbers) and the dynamo model output.
The nonlinear mean-field $\alpha\,\Omega$ dynamo produces the large-scale magnetic flux which is redistributed by negative effective magnetic pressure instability (NEMPI) producing sunspots and active regions.
The nonlinear mean-field dynamo model includes algebraic nonlinearity
(caused by the feedback of the growing magnetic field on the plasma motion)
and dynamic nonlinearities
(related to the dynamics of the magnetic helicity of small-scale magnetic field).
We compare the forecast errors with a horizon of 1, 6, 12 and 18 months, for different forecast methods,
with the same corrections on the current monthly observations.
Our forecast is in good agreement with the observed solar activity, the forecast error is almost stably small over short-medium ranges of forecasting windows.
Despite a strong level of chaotic component in the solar magnetic activity we present quantitative evidence that the solar activity on a short range can be stably well predicted, by the joint use of the physically based model with the neural network. This result may have an immediate practical implementation for predictions of various phenomena of solar activity and other astrophysical processes, so may be of interest to a broad community.

\end{abstract}

\section{Introduction}
\label{sect1}

Predictions of the solar activity is a crucial problem related to a fundamental solar physics
that have important applications. Various methods
including the mean-field dynamo models have been applied
to predict the solar activity
\citep[see, e.g.,][]{DG06,CCJ07,KA07,BT07,OS08,DJD09,KK11,PE12,TL15,KI16,U17,SKR18}.
Besides that there are other numerous methods to predict the solar activity using various sets of data as precursors or signatures of the forthcoming solar activity level.
Most of solar activity predictions are focused on the magnitude of the nearest solar cycle maximum, or the minimum, or even the magnitude of the further several cycles.
However, in spite of the numerous studies, improvement of the solar activity forecast is still a subject of active discussions.

Our approach is aimed for prediction of the 13-monthly running average
mean solar sunspot number.
We challenge our method for a predictions that is shifted a month ahead (or  6, 12, and 18 months ahead) with respect to 
the last available 13-month running mean average (i.e., with correction by up-to-date observations).
 Our method has been checked in real time with available observations over the last six years.
 The results and statistical data on the forecast since 2021 are available
 on the web\footnote{https://github.com/rodionstepanov/SolarActivityPrediction}.

In our approach to the forecast of solar activity, we take into account
the following:
\begin{itemize}
\item{
We use a nonlinear dynamo model \citep{KSR16,SKR18} that is able to reproduce properties and long-term dynamics of the mean magnetic field up to several hundreds of solar cycles.
This model describes the main cyclic oscillations of the large-scale magnetic field with an intrinsic chaotic behaviour caused by the dynamics of the magnetic helicity of small-scale magnetic fields.
The mean-field dynamo produces the large-scale magnetic flux \citep[see, e.g.,][]{M78,P79,KR80,ZRS83,MD19,RI21}.}
\item{
The redistribution of the large-scale magnetic flux by the large-scale negative effective magnetic pressure instability (NEMPI)  results in formation of sunspots and active regions \citep{KRR89,KRR90}.
This instability has been investigated theoretically using various analytical approaches \citep{KR94,KMR96,RK07}
and detected in direct numerical simulations in various setups \citep[see, e.g.,][]{BKR11,BRK16,WKR13,WKR16}.
NEMPI has a threshold in the magnitude of the mean magnetic field.
In the process triggered by this instability, no new large-scale magnetic flux is produced in contrast to the dynamo process.}
\item{
There are three characteristic times of interest concerning the solar activity prediction:
\newline
(i) long-term evolution of the mean magnetic field associated with the effect of magnetic helicity relaxation \citep{KR82,KRR95};
\newline
(ii) turbulent diffusion 
of the dynamo generated large-scale magnetic fields
at the scale of the solar convective zone depth.
These processes are described by the non-linear $\alpha\Omega$ dynamo model which takes into account evolution of small-scale magnetic helicity.
\newline
(iii) Short-term evolution at the scales of super-granulation related to formation of sunspots and active regions. This is entirely connected with NEMPI process. 
Decay of sunspots and active regions is included in this process, too.}
\item{
To predict the solar activity on a short time-scale, we combine the two approaches of the numerical solution of the nonlinear mean-field dynamo equations
and the artificial neural network.
The latter serves here as a correction scheme for the forecast,
which uses the currently available observational data (the 13 month running average of the observed
solar sunspot numbers time series) and the dynamo model output.}
\end{itemize}

The purpose of this paper is to comprehensively analyse the results of several years of practical forecasting of solar activity by the novel method \citep{SKR18}
which combines the solar dynamo model with magnetic helicity evolution and the neural network, estimate the errors of forecasting and demonstrate its capabilities through newly observable data in comparisons with other forecasting methods.

Despite a strong level of chaotic component in the solar magnetic activity we will present quantitative evidence that the solar activity on a short range can be reliably predicted, and it would give a positive example of the joint use of the physically based model plus the neural network.

\section{Comparison of the solar activity forecast methods.}
\label{sec2}

It is not our goal to review all papers dealing with the forecast of solar activity. We only want to point out 
inherent shortcomings of the basic forecasting methods
that we tried to avoid in this article.

The most common method is to predict the sunspot number based on the available series of observations for 24 cycles, i.e., about 270 years. The advantage of this method is the length of the series and the absence of any additional assumptions. In fact, this is a statistical extrapolation method for long series with some additional statistical details taken into account, such as 
the  growth rate of magnetic field in the cycle,
the relationship between the length and height of the cycle, and the observed relationship between the heights of two successive cycles of the Gnevyshev-Ohl type. Unfortunately, all existing methods are extremely unstable and do not give reliable results. Sometimes, the forecast comes true, sometimes it doesn't. However, even if the forecast does come true, it does not teach us anything, because it does not rely on understanding the underlying physics and the basic generation mechanisms. The next time, the same method may give a completely wrong result. Nevertheless, the above-mentioned method known as "forecasting by current measurements" proved useful for relatively short time intervals. In this method, the forecast is continuously refined by direct extrapolation or by more sophisticated methods, such as neural networks. The most widely known prediction methods for a term of several months are the McNish-Lincoln method \citep{ML49}, the standard method, and the combined method. The former forecasts are published by the National Geographic Data Center and the two latter ones, by the Solar Influence Data Analysis Center 
(for more details see \cite{PL12}).

\cite{Dmetal00} studied the relationship between the height of the cycle and duration of the cycle and its different phases. Since the duration of the cycle is not always clearly defined, the concept of the cycle current length determined by the autocorrelation function 
has been introduced in this method.
A statistically significant relationship between the maximum sunspot number and the duration of the cycle growth phase was confirmed. Besides that, a high correlation (of the order of 0.95) was shown to exist between the maximum amplitude of the cycle and the time derivative of the monthly Wolf numbers at the very beginning of the cycle growth phase.

\cite{KA07} performed a spectral analysis of the sunspot number time series to detect a periodicity using the maximum entropy method. 
He also used
the obtained periodicity to estimate the amplitude of Cycle 25 with the mean value of 119 and the maximum in 2022-2023. 

In 2009, De Jager \& Duhau showed that solar activity was changing from a Grand Maximum to a different regime \citep{DJD09}. The transition started in 2000 and was expected to last until the maximum of Cycle 24. After that, a short Grand Minimum similar to the Dalton one had to begin. 
This transition from moderate to low activity
was supposed to last for at least 60–100 years.

The precursor method can take into account many other indices that may be indicative of still unknown relationships between various characteristics of the sunspot formation activity. One can consider radio emission at different wavelengths, since it is associated with different objects on the Sun (faculae, spots, corona), characteristics of coronal holes, sector structure of the interplanetary magnetic field, the coronal green line brightness etc. 

In particular, \cite{Badetal01} predicted a strong decrease in the height of Cycles 23 and 24. \cite{TL09} examined the correlation of various characteristics of the epoch of minimum activity (in particular, the dipole-octupole index, the area and mean latitude of the field of dominant polarity in each hemisphere, the activity in polar faculae and K Ca II bright dots, and the intensity of the 5303 Å coronal emission line) with the amplitude of the forthcoming sunspot cycle. \cite{OS08} applied different prediction methods to Cycle 24. They proposed three forecasting indices: the polar field intensity, the mean field on the source surface, and the geomagnetic disturbance recurrence index. As a rule, the forecast based on the polar field and extrapolation of local fields 
predicts a lower height for Cycle 24 in comparison with that for Cycle 23.
Later, \cite{obrshel09} showed that the intensity of the polar magnetic field was steadily decreasing over the past three solar cycles. It is due to the fact that the increase in the dipole magnetic moment observed from 1915 to 1976 
was replaced by a decrease. 
At the same time, the medium-scale magnetic fields (e.g., the fields of isolated coronal holes) were unusually strong. The large effective contribution of the medium-scale fields to the total energy of the large-scale fields is also confirmed by calculations of the effective multipolarity index. The aa index at the cycle minima correlates with the height of the subsequent maxima. All this was interpreted as precursors of several low or medium cycles.

\cite{Upthat23} analyzed sunspot data for three years after the minimum of Cycle 24 and concluded that the sunspot number at the maximum of Cycle 25 should be 135 ± 10, i.e., slightly higher than in Cycle 24 (116.4). They also considered forecasts of Cycle 25 based on a number of precursors. The geomagnetic precursor (aa-index) suggested that Cycle 25 would be slightly higher than Cycle 24, with a maximum sunspot number of 132 ± 8. According to magnetic precursors (the polar field intensity and the axial dipole moment at the minimum), Cycle 25 was expected to be similar to Cycle 24, with a maximum sunspot number of 120 ± 10 or 114 ± 15. Some forecasts are based on geomagnetic field variations (aa-index) during two years before and two years after the sunspot minimum. When this method was proposed in 
the mid-50s of the last century
\citep{Ohl66, Ohl68, Ohl76, Ohl2-79}, it was purely empirical. But now, it is clear that geomagnetic activity during the minimum is closely related to large-scale magnetic fields on the Sun, namely the ones 
that form the sunspots
of the upcoming cycle. A disadvantage of this method is the need to use smoothed data for two years after the minimum. Since the length of the raise branch is often very short, the forecast lead time turns out to be no more than 2 years. \cite{Obr95} proposed an improvement, which increased the lead time of the forecast. It is interesting to note that this work was the first to point out a possible violation of the Gnevyshev-Ohl rule in the pair of Cycles 22-23, which was confirmed 10 years later.

Another method is based on direct use of large-scale magnetic field measurements. According to the generally accepted theory, the field of local regions arises from the poloidal magnetic field. A proxy of the latter can be the field in the polar regions, which is measured directly by magnetographs. It is true that these measurements are not too precise, since the field at the pole is mainly perpendicular to the line of sight, and, therefore, the magnetographs give a large error. Nevertheless, the data available show a high correlation between the magnitude of the polar field and the number of sunspots \citep{Biswasetal23}.

However, the polar field reaches its maximum in the vicinity of the sunspot minimum and not simultaneously in both hemispheres. Therefore, one has to wait about a year after the sunspot minimum to take reliable measurements. The main problem is that the forecast of the polar field itself is not a fully solved problem. Recently, forecasts of the polar field with increased lead time (i.e., several years before the onset of the minimum) have appeared. Thus, the amplitude of a solar cycle can be predicted as early as a few years after the field reversal in the previous cycle, thereby shifting the solar cycle forecast to much earlier times than usual \citep{Kumaretal21, Kumaretal22, Pishvas23}. The forecast made with such an increased lead time is generally consistent with others, including that of the Royal  Observatory of Belgium (ROB) service (https://sidc.be/SILSO/forecasts,2023) and yields values of about 135.

Finally, the most promising method is the direct application of the dynamo mechanism with appropriately selected parameters. The difficulty here is in choosing the optimal parameters. There are so many of them that it is simply impossible to go through them all directly. In addition, the question of how one cycle differs from another has not yet been finally resolved. In fact, it is necessary to select the contribution of the stochastic component for each cycle separately. To what extent the solar dynamo is determined by stochastic or deterministic processes is still unclear \citep{Minetal2002}.

Furthermore, the existing dynamo models are dealing with a mean-field dynamo; so, the output of any theory is the field structure, not the sunspot number that should be predicted. \cite{BT07} generally concluded that models based on a mean-field dynamo cannot be used to predict the solar cycle: "Given the inherent uncertainties in determining the transport coefficients and nonlinear responses for mean-field models, we argue that this makes it impossible to predict the solar cycle using the output from such models".

\cite{Kitkos08, KK11} used data assimilation methods to solve this problem. These methods combine observational data and models to estimate most accurately the physical properties that cannot be observed directly. The methods are able to provide a forecast of the future state of the system. It was shown that the ensemble Kalman filter (EnKF) method could be used to assimilate sunspot data into a nonlinear 
mean-field $\alpha\Omega$ dynamo model 
taking into account the dynamics of turbulent magnetic helicity. The forecast of Cycle 24 given in the old V1 system proved to be quite successful (approximately 60 units). \cite{KI16} assumed that Cycle 25 will be slightly lower than Cycle 24 and its maximum will take place in 2024.

As a rule, the physically based forecasting methods involve a scheme for calculating the polar field as a precursor of the following cycle. Such predictions use the flux transport dynamo (FTD) models, surface flux transport (SFT) models, or their combination. Thus, calculations by 
\cite{Diketal06} and \cite{Chetal07} 
based on similar initial assumptions, yield forecasts for Cycle 24 that differ by more than a factor of two. The convergence of the forecasts for 
Cycle 25 
obtained using the physically based models \citep{Upthat18, Jiangetal18, Bhownan18,  Labetal19} is somewhat better (110, 125, 118, and 89, respectively). The mean value according to calculations of  \cite{nandy21} is 110.5 $\pm13.5$ SSN, i.e., slightly lower than in Cycle 24 (116.4).

The recent physically based models created using state-of-the-art data also predict values somewhat higher than those of Cycle 24. \cite{Guoetal21} argue that Cycle 25 will be about 10\% higher than Cycle 24, with an amplitude of 126 (International Sunspot Number, version 2.0). \cite{Jiangetal23} examined seven models, two of which are based on the Flux Transport Dynamo, four, on the surface flux transport (SFT), and one is a mixed model. All of them strongly depend on the input data. Generally speaking, the physically based solar forecast is still in its early stage. It is an effective way to verify our understanding of the solar cycle. This work confirms that the polar field determines the subsequent cycle and that the Babcock–Leighton mechanism seems preferable. A similar conclusion 
was drawn by \cite{Bhowetal23}.

The main trends described above underlie all the numerous forecasts that appear at the beginning of each cycle and have so far given rather unconvincing results (e.g., see a detailed review by \cite{nandy21}). We will not cite and analyze all of them here. Nandi cites in his work 77 forecasts for Cycle 24 ranging from 60 to 250 SSN with the average of 163.1 $\pm42.2$ (the real value from smoothed annual means in April 2014 was 116.4) and 34 forecasts for Cycle 25 ranging from 50 to 220 SSN with a mean of 136.2$\pm41.6$.

Unfortunately, we must acknowledge that these forecasts, despite their good convergence, are not precise. In May 2023, the smoothed SSN number reached 123.9 and continued to grow steadily. Most likely, Cycle 25 should be at least 10\% higher than Cycle 24 \citep{ObrSS23, Obretal23a, Obretal23b}. But this forecast also turned out to be underestimated. The strong increase in activity in the second half of 2024 changed the course of the cycle, and the smoothed value for March 2024 was 141.3. At present, according to the ROB forecast service ((https://sidc.be/SILSO/forecasts), the maximum is expected sometime between July and November 2024 and its height is about 160.

In this paper, we attempt to combine the advantages of all the methods and eliminate the above-mentioned disadvantages. Our general idea is that the mean field dynamo 
creates a flux of the toroidal mean magnetic field,
which by some mechanism breaks up 
into separate magnetic flux tubes.
In this case, if the transverse size of the tube is small enough, we can expect that the field in the tubes will be noticeably stronger than the original toroidal mean field, and this will lead to the appearance of a photometrically detectable sunspot. In fact, this is no longer a dynamo, since no new magnetic flux is generated. This is 
another mechanism,
which transforms the mean 
magnetic field generated at the dynamo stage
into a set of sunspots. 
In this case, the equations involve stochastic parameters that account for cycles of different heights.

The model contains 9 parameters, but only 5 parameters are important.
The parameters are selected by comparing the calculations with the full set of sunspot data since 1750. After processing the model results, we obtain a time series of model sunspot numbers, which, although not reflecting all observed solar cycles exactly, nevertheless shows a very good correlation with them (above 85\%), including the amplitude and shape of these cycles. Thus, the model gives us approximate future cycles that can be brought to a real forecast by using the neural network method based on assimilation of current observations of the smoothed series of monthly mean sunspot numbers.

The model was created and tested in several stages. The data for the last four activity cycles were divided into training, validation and test samples. At the last stage, it is possible to move on to the forecast of smoothed sunspot numbers based on current measurements. Our task was not to give a forecast of the height of the upcoming cycle, but since the expected maximum phase of the 25th cycle is currently included in our forecast horizon, we provide the height and date of the maximum within the current forecast.

\section{Dynamo model and sunspot formation}
\label{sec3}

We adopt the following model  \citep{KSR16,KSR20,KRS23,SKR18} related to
the axisymmetric mean-field $\alpha \, \Omega$ dynamo,
which produces the large-scale magnetic flux that can be redistributed to form sunspots by NEMPI.
The axisymmetric large-scale magnetic field is written as  $ \meanBB = \meanB_\varphi
{\bm e}_{\varphi} + \bec{\nabla} {\bf \times} (\meanA {\bm e}_{\varphi})$,
where $r, \theta, \varphi$ are the spherical coordinates and ${\bm e}_{\varphi}$ is the unit vector.
The $\alpha \, \Omega$ dynamo equations as in the framework of the no-$r$ model are given by:
\begin{eqnarray}
{\partial \meanB_\varphi \over \partial t} &=& D \, \sin \theta {\partial \over \partial \theta}\meanA
+ \left({\partial^2 \over \partial \theta^2} - \mu^2 \right)\meanB_\varphi ,
\label{S1}
\end{eqnarray}
\begin{eqnarray}
{\partial \meanA \over \partial t} &=& \alpha \meanB_\varphi + \left({\partial^2
\over \partial \theta^2} - \mu^2 \right) \meanA ,
\label{S2}
\end{eqnarray}
where the coordinate $r$ is measured in the units of the solar radius $R_\odot$, the time $t$
is measured in the units of turbulent magnetic diffusion time $R_\odot^2 / \eta_{_{T}}$;
the toroidal field, $\meanB_\varphi(t,r,\theta)$, is measured in the units of $B_\ast$, where
$B_\ast = \xi \,\, \meanB_{\rm eq}$ with $\xi= \ell_0/\sqrt{2}R_\odot$
and $\meanB_{\rm eq} = u_0 \, \sqrt{4 \pi \meanrho_\ast}$.
The  magnetic potential, $\meanA(t,r,\theta)$, of the poloidal field is measured in the units of
$R_{\alpha} R_\odot B_\ast$, where $R_{\alpha} = \alpha_\ast R_\odot / \eta_{_{T}}$,
the fluid density $\meanrho(r,\theta)$ is measured in the units $\meanrho_\ast$,
the differential rotation $\delta\Omega$ is measured in units of the maximal value
of the angular velocity $\Omega$ and
the $\alpha$ effect is measured in units of the maximum value of the
kinetic $ \alpha $ effect, $\alpha_\ast$.
The integral scale of the turbulent motions
$\ell_0$ and the characteristic turbulent velocity $u_0$ at the scale $\ell_0$ are measured in units of their
maximum values in the convective zone.
The turbulent magnetic diffusion coefficient is $\eta_{_{T}}=\ell_0 \, u_0 / 3$.
The dynamo number is defined as $D = R_\alpha
R_\omega$, where $R_\omega = (\delta \Omega) \, R_\odot^2 / \eta_{_{T}}$.

The turbulent diffusion of the mean magnetic field
in the radial direction in this no-$r$ model is described by equations~(\ref{S1}) and~(\ref{S2})
by $-\mu^2 \meanB_\varphi$ and $-\mu^2\meanA$ \citep{KKMR03}.
The differential rotation is determined by factor
$G =\partial \Omega / \partial r$, which is taken zero in the vicinity of the equator 
\begin{equation}
G = \left\{ {\begin{array}{*{20}{c}}
{0,}&{\pi/2 - \varepsilon< \theta  < \pi/2 + \varepsilon}\\
{1,}&{else}
\end{array}} \right. .
\label{eq_G}
\end{equation}
The parameter $\mu$ is defined through
$\int_{2/3}^{1} (\partial^2 \meanB_\varphi / \partial r^2) \,dr = - (\mu^2/3) \meanB_\varphi$.
The value $\mu=3$ describes a convective zone with a thickness of 1/3 of the solar radius.

The total $\alpha$ effect,
\begin{eqnarray}
\alpha = \chi_{_{\rm K}} \Phi_{_{\rm K}}(\meanB) + \sigma_\rho \chi_{_{\rm M}} \Phi_{_{\rm M}}(\meanB) ,
\label{S3}
\end{eqnarray}
is the sum of the kinetic and the magnetic $\alpha$ effects,
where $\chi_{_{\rm K}} = - (\tau_0 /3) \, \langle\bec{u}\cdot(\bec{\nabla}
{\bf \times} \bec{u})\rangle$ is determined by kinetic helicity $\langle{\bm u} \cdot (\bec{\nabla} {\bf \times} {\bm u})\rangle$
and  
$\chi_{_{\rm M}} = (\tau_0 / 12 \pi \meanrho_\ast)\,\langle{\bm b} \cdot (\bec{\nabla} {\bf \times} {\bm b})\rangle$ 
is determined by current helicity of the fluctuation field 
$\chi_{\rm c}=\langle{\bm b} \cdot (\bec{\nabla} {\bf \times} {\bm b})\rangle$
.
 \citep{FPL75,PFL76},
and $\chi_{_{\rm K}}$ and $\chi_{_{\rm M}}$ are measured in units of maximum value of the $\alpha$-effect, $\alpha_\ast$.
Here $\tau_0$ is the correlation time of the turbulent velocity field,
${\bm u}$ and ${\bm b}$ are velocity and magnetic fluctuations,
and $\sigma_\rho = \int_{2/3}^{1} (\meanrho(r)/ \meanrho_\ast)^{-1} \, dr$. We adopted the profile of kinetic $\alpha$-effect in the form $$\chi_{_{\rm K}}(\theta) = \frac{{3\sqrt 3 }}{{2\cos ({\pi/6})}} \cos \left( \theta \right) \cdot \left[ {1 - {{\left( {\frac{{\cos \left( \theta \right)}}{{\cos \left( {{\pi/6}} \right)}}} \right)}^2}} \right].$$
The quenching functions $\Phi_{_{\rm K}}(\meanB)$ and $\Phi_{_{\rm M}}(\meanB)$ in
Eq.~(\ref{S3}) describe algebraic nonlinearity and are
given by \cite{FBC99} and \cite{RK00,RK04}:
\begin{eqnarray}
\Phi_{_{\rm K}}(\meanB) = {1 \over 7} \left[4 \Phi_{_{\rm M}}(\meanB) + 3 \Phi_{_{\rm B}}(\meanB)\right] ,
\label{S4}
\end{eqnarray}
and
\begin{eqnarray}
\Phi_{_{\rm M}}(\meanB) &=& {3\over \xi^2\meanB^2} \, \left[1 - {\arctan (\xi \, \meanB) \over
\xi \, \meanB} \right] ,
\label{S5} \\
\Phi_{_{\rm B}}(\meanB) &=& 1 - 2 \xi^2 \meanB^{2} + 2\xi^4 \meanB^{4} \ln \left[1 + (\xi^2\meanB^2)^{-1}\right] ,
\label{S6}
\end{eqnarray}
where
the mean magnetic field is given by
\begin{eqnarray}
\meanB^2 = \meanB_\varphi^2 + R_\alpha^2 \left[\mu^2 \meanA^2 + \left( {\partial \meanA \over \partial \theta}\right)^2\right] .
\label{S7}
\end{eqnarray}
The densities of the kinetic and current helicities, and quenching functions are associated with a middle part of the convective zone.
The parameter $\sigma_\rho > 1$ is a free parameter.

The function $\chi_{_{\rm M}}(\meanBB)$ describes dynamic nonlinearity, that is determined by the non-dimensional differential equation on current helicity of the fluctuation field
 $\chi_{\rm c}=\langle{\bm b} \cdot (\bec{\nabla} {\bf \times} {\bm b})\rangle$:
\begin{eqnarray}
&& {\partial \chi_{\rm c} \over \partial t} + \left(\tau_\chi^{-1} + \kappa_{_{T}}
\mu^2\right)\chi_{\rm c} = 2\left({\partial \meanA \over \partial \theta} {\partial \meanB_\varphi \over \partial \theta} + \mu^2 \meanA \, \meanB_\varphi\right)
\nonumber\\
&& \quad -  \alpha \, \meanB^2 - {\partial \over \partial \theta} \left(\meanB_\varphi {\partial \meanA \over \partial \theta} - \kappa_{_{T}} {\partial \chi^c \over \partial \theta} \right) ,
\label{S8}
\end{eqnarray}
where ${\bm F}_\chi = -\kappa_{_{T}} \bec{\nabla} \chi_{\rm c}$
is the turbulent diffusion flux of the magnetic helicity density of small-scale fields
that determines its transport \citep[see, e.g.,][]{KR82,KR99,KR22,KMR00,KMR02,BF00,BS05,GS23}
and $\kappa_{_{T}}$ is the coefficient of the turbulent diffusion of magnetic helicity.
In equation~(\ref{S8}), the time $\tau_\chi = \ell^2 / \eta$ is the relaxation time of magnetic helicity.
The average value of $\tau_\chi^{-1}$ is given by the estimation 
\begin{eqnarray}
\tau_\chi^{-1} = H_\ast^{-1} \int_{r_{\rm c}}^{1} \tilde \tau_\chi^{-1}(r) \,d r \sim {H_\ell \, R_\ast^2 \,
\eta \over H_\ast \, \ell^2 \, \eta_{_{T}}} ,
\label{S9}
\end{eqnarray}
where $H_\ast$ is the depth of the convective zone,
$H_\ell$ is the characteristic scale of variations $\ell_0$, and
$\tilde \tau_\chi(r) = (\eta_{_{T}} / R_\ast^{2}) (\ell_0^2 / \eta)$ is the non-dimensional
relaxation time of the density of the magnetic helicity. The values
$H_\ell , \, \eta, \, \ell_0$ in equation~(\ref{S9}) are associated
with the upper part of the convective zone.

\begin{figure}
\centering
\includegraphics[width=12.0 cm]{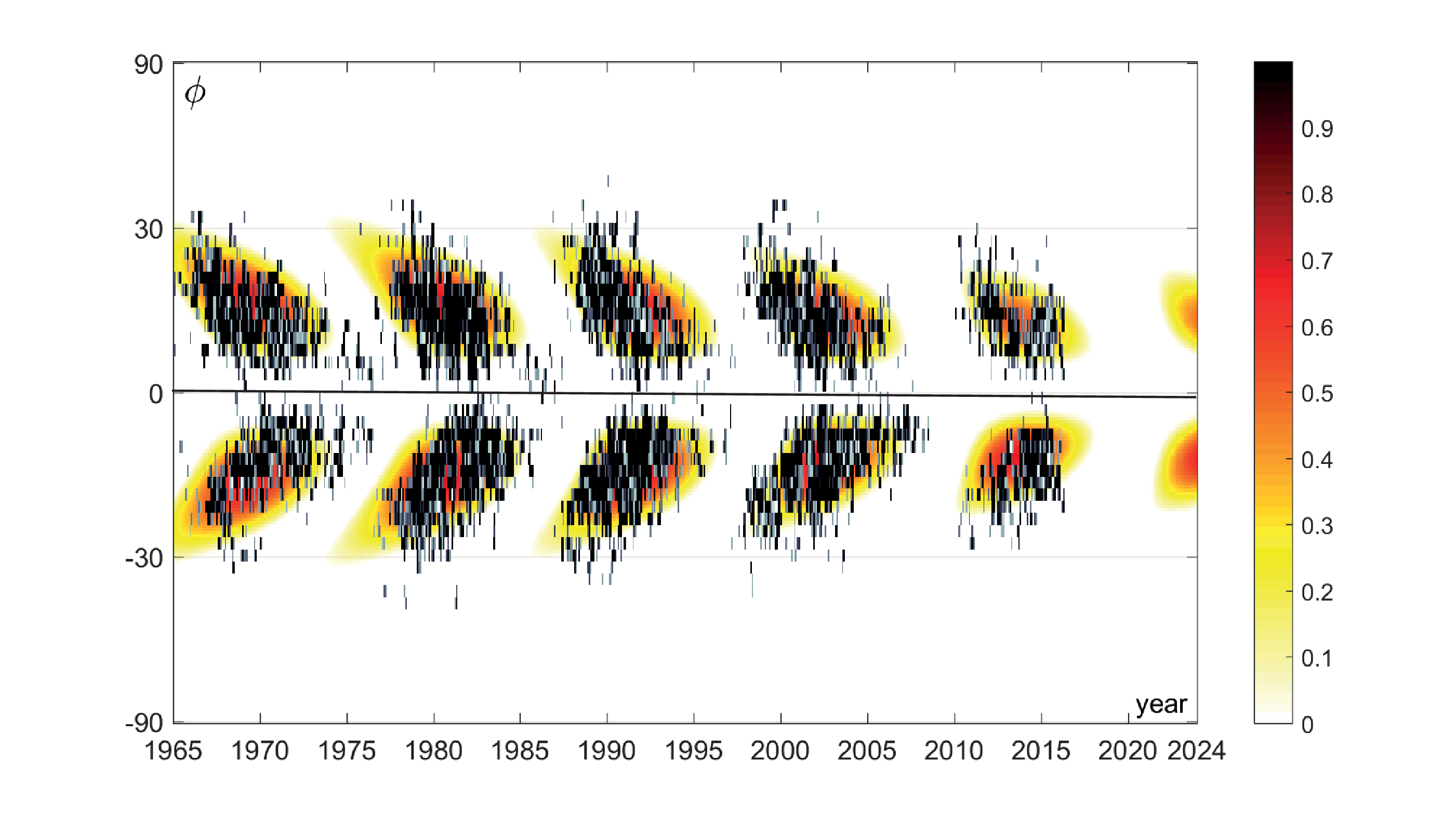}
\caption{\label{Fig1}
The butterfly diagram of the  solar sunspot number variation rate
$2 \pi \, \sin \theta \,  I(t,\theta)$ obtained using the dynamo model (colour) and
the real monthly observational data (black).
}
\end{figure}

An important possible mechanism of sunspot formation is NEMPI.
For post-processing of the model solution
we amplify
global dipole magnetic field by reflection transformation $0.5\left[ {{\meanB }\left( {\theta ,t} \right) - {\meanB }\left( {\pi  - \theta ,t} \right)} \right] +  0.05\left[ {{\meanB }\left( {\theta ,t} \right) + {\meanB }\left( {\pi  - \theta ,t} \right)} \right] $.
Based on the ideas of NEMPI, we
derive a budget equation for the surface density of the solar sunspot number
\citep{KSR16,SKR18}:
\begin{eqnarray}
{\partial \tilde W \over \partial t} = I(t,\theta) - {\tilde W \over \tau_s(\meanB)} ,
\label{S10}
\end{eqnarray}
which includes the rate of production of the surface density of the  solar sunspot number, $\tilde W(t,\theta)$,
due to the formation of sunspots:
\begin{eqnarray}
I(t,\theta) = {|\gamma_{\rm inst}| |\meanB-\meanB_{\rm cr}| \over \Phi_s} \Theta(\meanB-\meanB_{\rm cr}) ,
\label{S11}
\end{eqnarray}
and the rate of decay of the surface density of the  solar sunspot number, $\tilde W / \tau_s(\meanB)$,
which mimic the decay of sunspots.
Here $\tau_s(\meanB)$ is the decay time of sunspots,
$\Theta(x)$ is the $\Theta$ function defined as $\Theta(x) = 1$ for $x>0$, and $\Theta(x) = 0$ for $x\leq 0$,
$\Phi_s$ is the magnetic flux inside a magnetic spot,
and $\gamma_{\rm inst}$ is the growth rate of NEMPI, given in Appendix 
\citep[see also,][]{RK07,BRK16}.
The solar sunspot number is defined as a surface integral:
$W = R_\odot^2 \, \int \tilde W(t,\theta) \sin \theta \, d\theta \, d\phi
= 2 \pi \,  R_\odot^2 \, \int \tau_s(\meanB) \,I(t,\theta) \sin \theta \,d\theta$.
To determine the function $\tau_s(\meanB)$, we take into account that
when the solar activity increases (decreases), the average life time of sunspots increases (decreases), so that $\tau_s(\meanB)$ is
$\tau_s(\meanB)=\tau_\ast \exp \left(C_s \, \partial \meanB/\partial t\right)$
with $C_s= 5.47 \times 10^{-4}$ and $\tau_\ast \, \gamma_{\rm inst} \sim 10$.
Here the non-dimensional rate of the mean magnetic field, $\partial \meanB/\partial t$,
is measured in units $\xi \meanB_{\rm eq} / t_{\rm td}$, and $t_{\rm td}$ is the turbulent magnetic diffusion time.
A particular form of function $\tau_s(\meanB)$ weakly affects the dynamics of solar sunspot numbers.
Equation~(\ref{S10}) provides the correspondence between the surface density of the total  sunspot number and dynamo generated magnetic fields. The equation has no stochastic ingredient except the mean magnetic field itself that is obtained by solution of equations~(\ref{S1}), (\ref{S2}), and (\ref{S8}).

\begin{figure}
\centering
\includegraphics[width=12.0 cm]{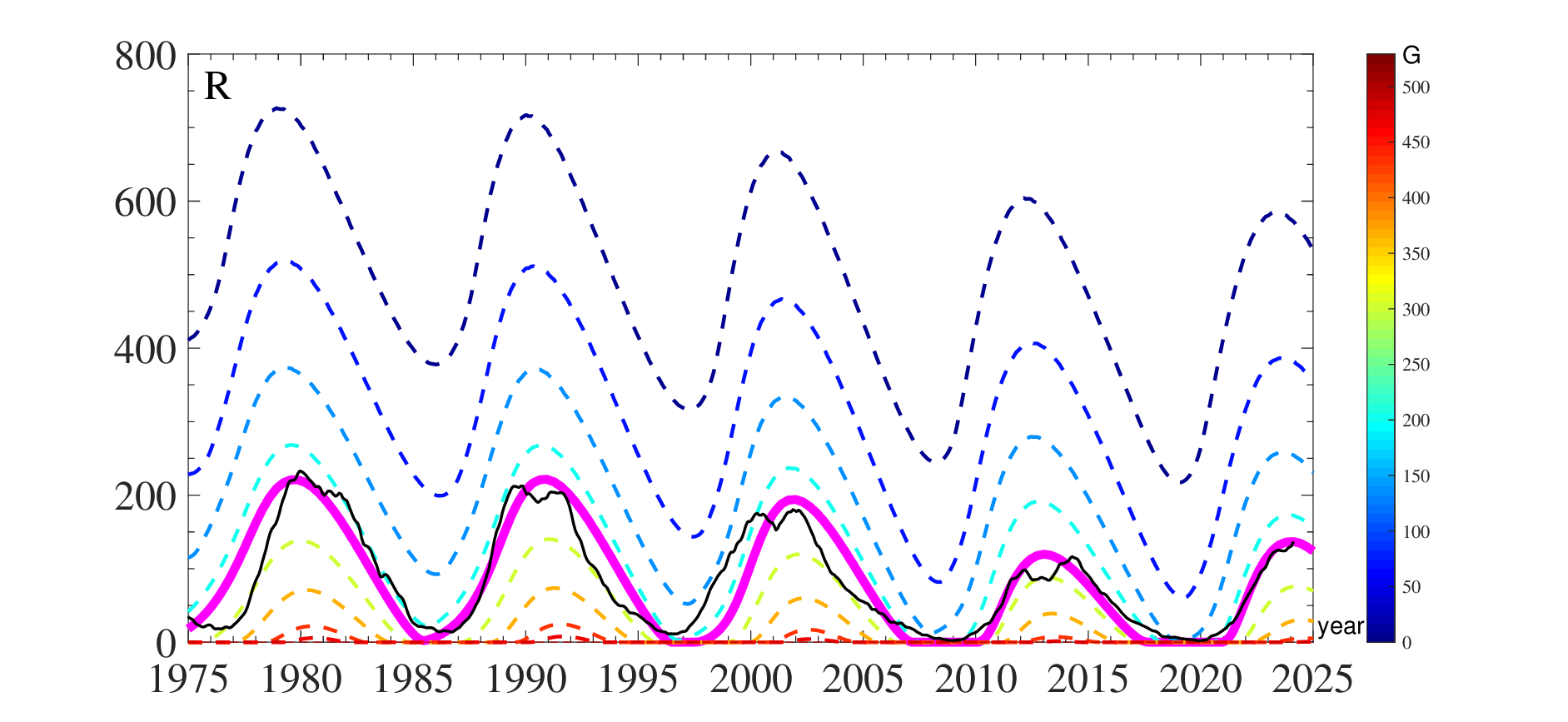}
\caption{\label{Fig2}
The height of solar cycles computed with various threshold values $\meanB_{\rm cr}$ required for the excitation of NEMPI,
where the black line corresponds to the real monthly observational data and the solid magenta line corresponds to $\meanB_{\rm cr}=265$ G.}
\end{figure}

Equations~(\ref{S1}), (\ref{S2}), (\ref{S8}) and~(\ref{S10}) have been solved numerically.
We use MATLAB code, which solves initial-boundary value problems for systems of partial-differential equations that employs a second-order explicit finite differences scheme in space.
We use the spatial resolution of $203$ mesh points in co-latitude $\theta$
(this odd number provides mesh intervals below 1 degree). We choose $\varepsilon  = 2\pi/203$ that means clipping just one mesh point in $G$ near the equator.
The time grid in simulations varied between $6 \times 10^5$ and $18 \times 10^5$
time instants for a different set of initial parameters due to long transitional processes.

We apply the following initial conditions: $\meanB_\phi(t=0,\theta)=S_1 \sin\theta + S_2 \sin(2\theta)$ and $\meanA(t=0,\theta)=0$
corresponding to a combination of the dipole and the quadruple type of solutions.
The  boundary conditions are $\meanB_\phi(t,0)=\meanB_\phi(t,\pi)=0$; $\,\meanA(t,0)=\meanA(t,\pi)=0$, and
$\,\partial \chi^c(t,0)/\partial \theta=\partial \chi^c(t,\pi)/\partial \theta=0$.
We use the following values of the governing parameters: $D=-8450$, $\sigma_\rho=3$, 
$\kappa_{_{T}}=0.1$, $R_\alpha=2$, $T=6.3$, $S_1=0.051$, $S_2=0.95$.
We have used these parameters and initial conditions for various modelling of the solar and stellar activity
by the axisymmetric mean-field $\alpha \, \Omega$ dynamo \citep{KSR16,KSR20,KRS23,SKR18},
where the mechanism of the sunspot formation by NEMPI have been taken into account.
In addition, the parameter $\mu=3$ corresponds to the solar convective zone, while the parameter $\xi=0.3$ is used to compute $\tau_s(\overline{B})$.
This particular choice of model and post-processing parameters has been made as providing the best fit of model time series $W$ to the observational data of solar sunspot number v.2.0 for solar cycles 20-24 taken from the World Data Center SILSO, Royal Observatory of Belgium, Brussels.
Figure~\ref{Fig1} we shows the butterfly diagram of the  solar
sunspot number 
obtained from the dynamo model 
compared 
with the observational data.
We take into account that NEMPI has a threshold 
for the magnetic field%
.
The effect of the threshold is illustrated in Fig.~\ref{Fig2},
where we show the height of solar cycles computed with various cut-off values
for the excitation of NEMPI.

\section{Forecast of the solar activity}
\label{sect4}

Mean-field dynamo models are relevantly applicable on a time-scale that is larger than, say, one year,
and alone it cannot provide an accurate forecast of the solar activity on a time-scale of a few months.
To predict the solar activity on a short time-scale, we use a method based on a combination of the numerical solution of the nonlinear mean-field dynamo equations and the artificial neural network approach [see for details, \cite{SKR18}, and references therein].
To apply this approach, we initially used the original simulations of the solar sunspot number series $W_i^{\rm model}$ based on the dynamo model described in Sec.~\ref{sec3}, as the basis for the forecast, and as the exogenous input in the neural network scheme.
Another input is the data $W_i^{\rm obs}$ obtained from observations (the 13 month running average of the observed solar sunspot number time series).
To perform the forecast $W_i^{\rm forecast}$, we adopt a ''two-layer artificial neural network'', which is a recurrent dynamic nonlinear autoregressive network, with feedback connections enclosing two layers of the network, defined by the following equation:
\begin{eqnarray}
W_i^{\rm forecast} = {f_{\rm out}}\left[ {{{\bm{K}}_{2}}\, {f_{\rm hidden}}\left( {{{\bm{K}}_{1}}\, {\bm{w}} + {{\bm{c}}_1}} \right) + {{\bm{c}}_2}} \right],
\label{T2}
\end{eqnarray}
where $f_{\rm hidden}(x) = [1 + \exp(-x)]^{-1}$ is a function of a hidden layer of neurons, $f_{\rm out}(x)=x$,
${\bm{K}}_{1}$ is the weight matrix $24 \times 8$ of a hidden layer neurons,
${\bm{K}}_{2}$ is the weight matrix $1 \times 24$ of an outer layer neurons,
${{\bm{c}}_1}$ and ${{\bm{c}}_2}$ are the corresponding bias vectors,
$\bm{w}$ is the input vector $8 \times 1$ consisting of 4 prior observations
$\begin{array}{*{20}{c}} {W_{i - 1}^{{\rm{obs}}}}, &  \cdots \,, & {W_{i - 4}^{{\rm{obs}}}}  \\ \end{array}$ and 4 corresponding model estimations
$\begin{array}{*{20}{c}} {W_{i}^{{\rm{model}}}}, &  \cdots \,, &  {W_{i - 3}^{{\rm{model}}}}  \\ \end{array}$.

\begin{figure}
\centering
\includegraphics[width=12.0cm]{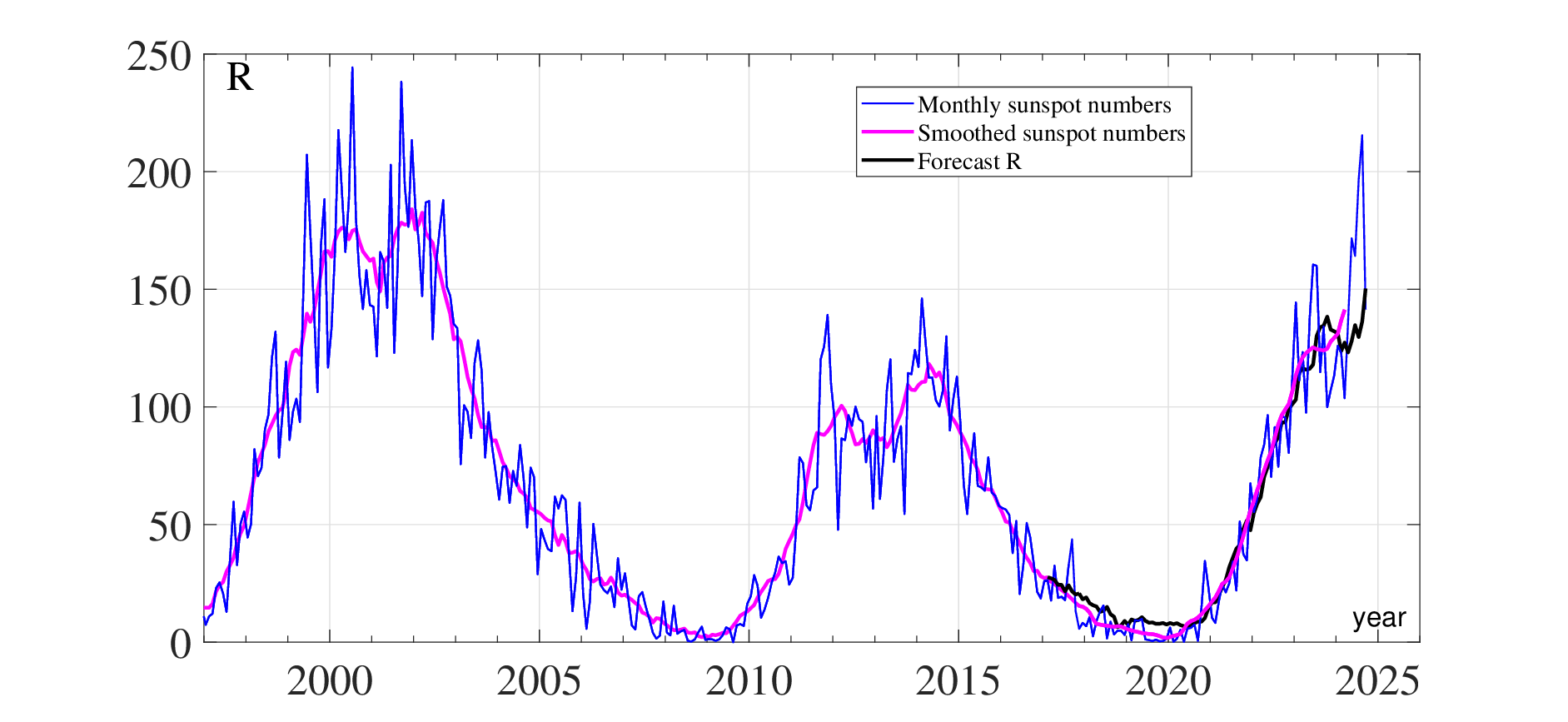}
\caption{\label{Fig3}
The one-month forecast of the solar activity (red line)  compared with the observed solar sunspot numbers running average over 13 months (blue line) and the test sample forecast (black line).}
\end{figure}

\begin{figure}
\centering
\includegraphics[width=15.0cm]{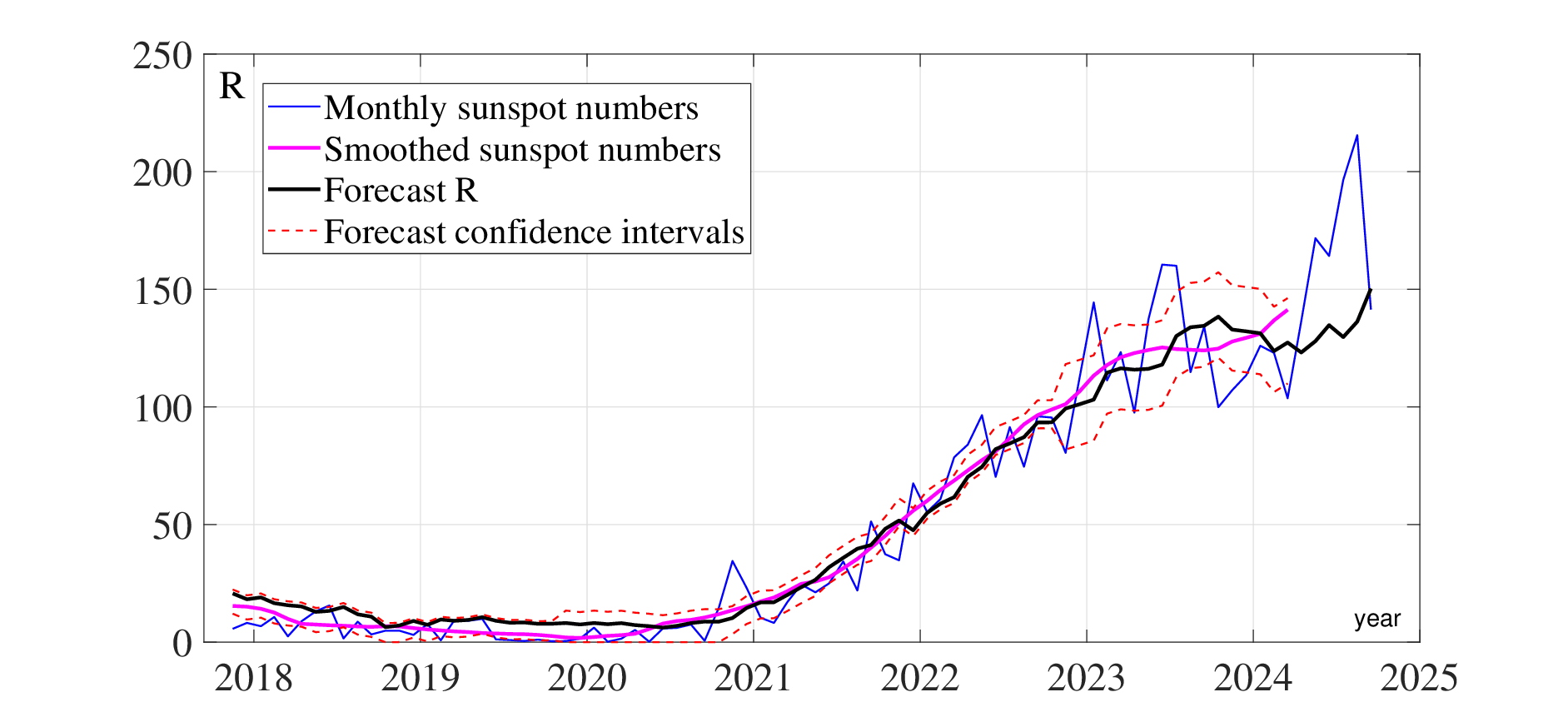}
\caption{\label{Fig4}
Results of forecasting the solar activity obtained by our approach.
}
\end{figure}

The learning procedure by Bayesian regularization back-propagation was based on epignose using the data of the  solar sunspot numbers from  cycles 20-21, while cycle 22 has been used for the validation process.
The input data of the  solar sunspot numbers for the neural network consist of two parts: the prior real observations
and the dynamo model estimations at the same instant.
The output of this neural network is the forecasted monthly solar sunspot number.
We do not use the artificial neural network for any type of optimisation or parameter estimation for the initial basic (physical) model, that has already carried out earlier \citep{KSR16}.
During the learning procedure of the artificial neural network,
we minimize the error between the forecast and the actual observations not at every instant separately but over an entire cycle.
The dynamo model output is used as an initial forecast, and
the artificial neural network is a correction scheme for the final forecast by means of
the currently available observational data and the dynamo model output.
The forecast confidence intervals of the one-month forecast of the solar activity compared with the observed solar sunspot numbers running average over 13 months and the test sample forecast are shown in Fig.~\ref{Fig3}.
The forecast of the solar activity is shown in Fig.~\ref{Fig4}, where we compare the results
of the one-month forecast of the solar activity based on the described method 
and the observed  solar sunspot numbers averaged by 13 month sliding window.

\begin{table}
\centering
\caption{\label{Table}
Comparison of the forecast errors with a horizon 1, 6, 12, and 18 months for different forecasting methods. Forecast error are calculated in the interval from Sept 1997 to May 2010 except the line marked 
$\ast$ which corresponds to
the interval from
Nov 2017 to 
October 2024
for which the data are available on 
https://github.com/rodionstepanov/SolarActivityPrediction
.
}
\begin{tabular}{|l|l|l|l|l|}
\hline
Method                                               & \multicolumn{4}{c|}{RMS}       \\ \cline{2-5}
                                                          & 1 m & 6 m & 12 m & 18 m \\
\hline
NARX (Nonlinear Autoregressive Exogenous Model) with corrections$\ast$     & 1.24 &  5.10  &  6.21  & 7.03 \\
\hline
NARX without corrections   & 1.54   & 8.09   & 10.32   & 17.92   \\
\hline
M\&L method   & 3.6    & 5.9    & 10.9    & 15.2    \\
\hline
M\&L method with KF   & 3.1    & 4.9    & 9.3     & 12.4    \\
\hline
Standard Method (SM)      & 3.4    & 6.1    & 12.3    & 17.6    \\
\hline
Standard Method with KF  & 2.9    & 5.3    & 11.3    & 16.7    \\
\hline
Combined Method (CM)              & 4.7    & 10.4   & 17.5    & 17.5    \\
\hline
Combined Method with KF         & 3.2    & 6.0    & 13.1    & 16.4 \\
\hline
\end{tabular}
\end{table}

Qualitative comparison of the forecast errors with a horizon of 1, 6, 12 and 18 months, for different forecasting  methods
is presented in Table~\ref{Table}.
This implies that the forecast is shifted a month ahead (or  6, 12, and 18 months ahead) in comparison
with the last available 13-month running mean (i.e., with correction by up-to-date observations).
The notations in Table~\ref{Table} are the following:
our method (Nonlinear Autoregressive Exogenous Model, NARX) with monthly corrections with current observations
and without the corrections (just using only the mean-field dynamo model),
the McNish–Lincoln method \citep{ML49} (M\&L) with and without Kalman filter (KF),
the standard method (SM), and the combined method (CM), see for details \cite{PL12}.
For qualitative assessment of accuracy of our forecast,
we use the data from Table~5 published by \cite{PL12} with the correction
of standard forecasting methods by means of data assimilation method such
as Kalman filter.
These data are chosen because the modern forecasts of the average monthly number of sunspots
presented on the website of the Royal Belgian Observatory\footnote{https://www.sidc.be/SILSO/home}
are being produced on their basis.
One can see that at the shorter interval the errors are somehow greater as the  statistical properties of the process are not stationary (compare the first two lines in Table~\ref{Table}).
Even though the forecasts are fundamentally different from ours, they use the same corrections every month based on the current observations.

\section{Discussions and conclusions}
\label{sect5}

The comparison of our forecast with the observed solar activity demonstrate good agreement
(see Fig.~\ref{Fig3}).
We compared our results of forecasting with those by other methods (see Table~\ref{Table}).
It is notable that our prediction error is almost stable over short and longer ranges of forecasting windows.

We would like to stress that the advantage of our method is the combination
of the numerical solution of the nonlinear mean-field dynamo equations and the artificial neural network.
The mean-field dynamo model alone while gives plausible magnitudes of the forthcoming solar cycle level,
is not able to produce correct details of the sunspot number dynamics and its timing over the phases of the solar cycle.
Using only the neural network without an account of the mean-field solution
provides reasonable agreement with available observations for just a few years 
because in this case there is no long-term memory in the magnetic field evolution.

The currently available data series is non-stationary and the duration of this time series is very short.
So, the scientifically meaningful forecast can use the 13-monthly running average.
This averaging smooths the most prominent typical noise of the signal.
This is possible because the typical correlating times of the dynamo process
is much longer than the statistical background turbulence noise signal.

The disadvantage of this approach is that the results of prediction cannot
be verified immediately as one has to wait several months for the observable values.
The significant advantage is that the range of forecasting can  easily be extended to 6, 12 or even 18 months,
which with correction by up-to-date observations has stably small forecasting error.

Despite a high level of chaotic component in the solar magnetic activity, we demonstrate that the solar activity on a short time scale (up to 1.5 years) can be predicted with a good accuracy using a physically based model of the solar activity and the neural network.
This result may have an immediate practical implementation for predictions of various characteristics of solar activity and other astrophysical processes so may be of interest to a broad community.

\begin{acknowledgments}
The work of NK, KK, NS and RS was supported by the Russian Science Foundation (grant 21-72-20067).
IR would like to thank support and hospitality of NORDITA, Stockholm University and KTH Royal Institute of Technology (during the programme ''Towards a comprehensive model of the galactic magnetic field")
and the Isaac Newton Institute for Mathematical Sciences, Cambridge University (during the programme "Anti-diffusive dynamics: from sub-cellular to astrophysical scales").
\end{acknowledgments}

\appendix
\section{Characteristics of NEMPI}

The growth rate of NEMPI is given by
\begin{eqnarray}
&& \gamma_{\rm inst} \approx 
\left[ {2V_{\rm A}^2 k_x^2\over {H_\rho^2 {k}^2+1/4}}
\left|{d P_{\rm eff} \over d \beta^2}\right| - {{ 4 H_\rho^2({\bm \Omega} \cdot {\bm k})^2+(\Omega \sin{\phi})^2}
\over {H_\rho^2 {k}^2+1/4}}\right]^{1/2}
\nonumber\\
&& \quad - \eta_{_{T}} \left({k}^2 + {1 \over (2 H_\rho)^{2}}\right) ,
\label{S12}
\end{eqnarray}
where $V_{\rm A}=\meanB/\sqrt{4 \pi \meanrho}$ is the mean Alfv\'en speed,
${\bm k}$ is the wave number, ${\bm \Omega}$ is the angular velocity, 
$\phi$ is the heliographic latitude,
$P_{\rm eff}=\half\left[1-q_{\rm p}(\beta)\right]\beta^2$ is the effective magnetic pressure,
the nonlinear function $q_{\rm p}(\beta)$ is the turbulence contribution to the mean magnetic pressure
and $\beta=\meanB/\meanB_{\rm eq}$.
NEMPI is excited in the upper part of the convective zone,
where the Coriolis number ${\rm Co}= 2 \Omega \, \tau_0$ is small.
This implies that the instability is excited $(\gamma_{\rm inst}>0$),
when the mean magnetic field is larger than a critical value,
$\meanB_{\rm cr}$ that is given by
\begin{eqnarray}
{\meanB_{\rm cr} \over \meanB_{\rm eq}} \approx 
{\ell_0 \over 50 H_\rho} \left[1 + \left({10 \, {\rm Co} \, H_\rho^2 \over \ell_0^2} \right)^2 \right]^{1/2} .
\label{S14}
\end{eqnarray}
For the upper part of the convective zone, $\meanB_{\rm cr} \geq \meanB_{\rm eq} / 50$ is small.
The characteristic time of the  solar sunspot number variations is of the order of
the characteristic time for excitation of the instability, $\gamma_{\rm inst}^{-1}$.

\bibliography{paper3forecast}{}
\bibliographystyle{aasjournal}

\end{document}